\documentclass[superscriptaddress,showpacs,twocolumn,prl]{revtex4} 

\usepackage{amsmath}
\usepackage{amssymb}
\usepackage{dsfont}
\usepackage{color}
\usepackage{hyperref}
\usepackage{graphicx}

\newcommand{\ket}[1]{|#1\rangle}

\begin{document}

\title{Elementary excitations of chiral Bose-Einstein condensates}
\author{M. J. Edmonds} 
\email{matthew.edmonds@ncl.ac.uk}
\affiliation{Joint Quantum Centre (JQC) Durham-Newcastle, School of Mathematics and Statistics, Newcastle University, Newcastle upon Tyne NE1 7RU, England, United Kingdom}
\affiliation{SUPA, Institute of Photonics and Quantum Sciences, Heriot-Watt University, Edinburgh, EH14 4AS, United Kingdom}
\author{M. Valiente}
\affiliation{SUPA, Institute of Photonics and Quantum Sciences, Heriot-Watt University, Edinburgh, EH14 4AS, United Kingdom}
\author{P. \"Ohberg}
\affiliation{SUPA, Institute of Photonics and Quantum Sciences, Heriot-Watt University, Edinburgh, EH14 4AS, United Kingdom}


\begin{abstract}
We study the collective modes of a Bose-Einstein condensate subject to an optically induced density-dependent gauge potential. The corresponding interacting gauge theory lacks Galilean invariance, yielding an exotic superfluid state. The nonlinear dynamics in the presence of a current nonlinearity and an external harmonic trap are found to give rise to dynamics which violate Kohn's theorem; where the frequency of the dipole mode strongly depends on the strength of the mass current in the gas. The linearised spectrum reveals how the centre of mass and shape oscillations are coupled, whereas in the strongly nonlinear regime the dynamics is irregular.
\end{abstract}

\maketitle

{\it Introduction.} 
Interacting, degenerate many-body systems formed of bosons or fermions form the building blocks with which we can understand a broad spectrum of  phenomena at ultracold temperatures \cite{bloch_2008}. 
Here, interest has focused on understanding the emergent properties of these highly controllable macroscopic systems. 
Over the last few years the ability to mimic the behavior of charged particles with cold atomic gases has lead to intense activity both theoretically and experimentally \cite{dalibard_2011,wu_2011,goldman_2013,zhou_2013}. 
These charge neutral atomic systems formed from cold atoms benefit from being highly experimentally malleable, a situation which has lead to the simulation of orbital magnetism \cite{lin_2009a,lin_2009b}, as well as more exotic scenarios such as spin-orbit coupling with both bosons \cite{lin_2011} and fermionic gases \cite{fermion_so,cheuk_2012}.

Principally, there are several methodologies which can be used to induce artificial gauge potentials in neutral atoms. 
One can stir the condensate with a laser, which creates a uniform magnetic flux across the atomic cloud \cite{madison_2000}. 
Alternatively, one can optically dress the condensate with Raman couplings, or schemes based on dark state dynamics \cite{dalibard_2011}. 
By optically coupling several of the internal states of the atoms, one can simulate more elaborate gauge theories with these systems \cite{ruseckas_2005,osterloh_2005}, such as spin-1 spin-orbit coupling \cite{lan_2014}, as well as `interacting' gauge potentials, where there is a nonlinear feedback between the gauge field and the matter field \cite{aglietti_1996,edmonds_2013a,greschner_2013,zheng_2014}.
By performing these optical manipulations, we expect the fundamental properties of the many body system to change, hence it is important to understand how the nature of the condensed state is altered by such changes.
Insight into the dynamical response of superfluid systems can be accomplished by means of studying the elementary excitations of the many-body ground state.

The study of the excitations of many particle bosonic quantum systems gives fundamental insight into the response of the system to small perturbations. Theoretical studies of the collective modes of condensates in artificial gauge potentials has shown how the excitation spectrum undergoes a Zeeman like shift in the presence of a Landau gauge potential \cite{murray_2007}. Analysis of spin-orbit condensates with equal Rashba and Dresselhaus couplings has shown the excistence of a rich excitation structure, including  topological phases \cite{martone_2012}, and an excitation structure that reflects both the spin and density character of these condensates \cite{li_2013}.     

In this paper we study the nonlinear dynamics and the elementary excitations of a trapped one-dimensional Bose-Einstein condensate coupled to a density-dependent gauge potential (see also \cite{zheng_2014} for a study of 2D and 3D effects). The system is interesting from a fundamental point of view, providing a scenario which is governed by an interacting gauge theory, and as such can be seen as emulating a quasi dynamical field theory with some unusual nonlinear dynamics.  The frequencies of the collective excitations of the gas are found to be strongly depending on the emerging current nonlinearity, where in the strongly coupled regime one sees a type of non-equilibrium steady state appearing. As an alternative description of the dynamics we provide a variational calculation which is able to capture the strongly coupled situation which involves strong currents.  A trial Gaussian solution is used along with the underlying Lagrangian density to derive newton-like equations of motion for the centre of mass coordinate and the width of the condensate.


{\it Origin of the nonlinear gauge potential.} Creating artificial gauge potentials with atomic systems has reached an impressive level of accomplishment, ranging from inducing quantised vortices \cite{lin_2009b}, spin-orbit coupling \cite{lin_2011} and even suggestions for simulating dynamical gauge fields in optical lattices \cite{banerjee_2012,tagliacozzo_2013,zohar_2013}. A simple and illustrative setup for creating static gauge potentials is based on a coupled two-level system where the coupling is allowed to be detuned and/or space dependent \cite{dalibard_2011} which is also similar in spirit to the first experiments on spin-orbit coupling, but now in the adiabatic regime where only one of the two eigenstates stemming from the coupled system is populated. The main concept for generalising this system to a nonlinear gauge potential is the observation that the gauge potential is proportional to the detuning of the coupling laser, hence if collisions induce a detuning then the resulting gauge potential will also depend on the density of the gas. A realistic Hamiltonian describing $N$ two-level trapped bosonic atoms with internal states $\ket{1}$ and $\ket{2}$ which includes the particle interactions as well as the coupling of the two internal states can be written as
\begin{equation}\label{ham1}
\hat{\mathcal{H}}=\left(\frac{\hat{\bf p}^2}{2m}+\frac{1}{2}m\omega_{t}^2x^2+\frac{1}{2}m\omega_{\bot}^{2}{\bf r}_{\bot}^{2}\right)\otimes\mathds{1}+\hat{H}_{lm}+\hat{\mathcal{V}},
\end{equation}
where
\begin{equation}\label{lm}
\hat{H}_{lm}=\frac{\hbar\Omega}{2}\left(\begin{array}{cc}0 & e^{-i\phi({\bf r})}\\ e^{i\phi({\bf r})} & 0\end{array}\right)
\end{equation}
describes the optical coupling between the states $\ket{1}$ and $\ket{2}$ whose strength is characterized by the two-photon Rabi frequency $\Omega$. To derive an equation of motion for the many particle system, we define the Hartree wave function as $|\Psi\rangle=\otimes^{N}_{l=1}|\chi^{(0)}_{l}\rangle$ where the state $|\chi^{(0)}_{l}\rangle$ is the single-particle wave function, which we take as one of the eigenstates of equation (\ref{lm}) above. The other quantities appearing in equation (\ref{ham1}) are the radial coordinate ${\bf r}_{\bot}$, while the axial and radial trap frequencies are defined as $\omega_{t}$ and $\omega_{\bot}$ respectively. The interactions are characterized by the matrix $\hat{\mathcal{V}}=(1/2)\rm{diag}[g_{11}\rho_{1}+g_{12}\rho_{2},g_{22}\rho_{2}+g_{12}\rho_{1}]$, where the population of state $i$ is denoted $\rho_i=|\Psi_i|^2$. We assume the Bose-Einstein condensate is sufficiently dilute such that we are allowed  to construct interacting dressed states using perturbation theory with the eigenstates of the light matter coupling from equation (\ref{lm}). Denoting the unperturbed dressed states as $\ket{\chi^{(0)}_{\pm}}=(\ket{1}\pm\exp\{i\phi({\bf r})\}\ket{2})/\sqrt{2}$, we obtain the perturbed dressed states as
\begin{equation}\label{ds}
\ket{\chi_{\pm}}=\ket{\chi^{(0)}_{\pm}}\pm\frac{g_{11}-g_{22}}{8\hbar\Omega}\rho_{\pm}\ket{\chi^{(0)}_{\mp}},
\end{equation}
while the eigenvalues are given by $g\rho_{\pm}\pm\hbar\Omega/2$, and the dressed scattering parameter is defined by $g=(g_{11}+g_{22}+2g_{12})/4$. 

To derive an interacting gauge theory, we define a state vector as $\ket{\xi}=\sum_{i=+,-}\Psi_i\ket{\chi_i}$, and project the adiabatic motion of the atoms into one of these two states. The effective Hamiltonian then becomes
\begin{equation}\label{eom}
\hat{H}_{\pm}=\frac{1}{2m}(\hat{\bf p}-{\bf A}_{\pm})^2+\frac{1}{2}m\omega_{t}^{2}x^2+\frac{1}{2}m\omega_{\bot}^{2}{\bf r}_{\bot}^{2}+\frac{g}{2}\rho_{\pm}.
\end{equation}
The density-dependent geometric phase that arises in equation (\ref{eom}) is given by ${\bf A}_{\pm}=i\hbar\langle\chi_{\pm}|\hat{\nabla}\chi_{\pm}\rangle$. There is also a scalar geometric phase, which in what follows acts only as an energy offset, and as such is dropped. From the definitions of the perturbed dressed states, equation (\ref{ds}), the leading order contribution to the vector potential is ${\bf A}_{\pm}={\bf A}^{(0)}\pm{\bf a}_{1}\rho_{\pm}({\bf r})$, where ${\bf A}^{(0)}=-\frac{\hbar}{2}\hat{\nabla}\phi({\bf r})$ defines the single-particle vector potential while ${\bf a}_{1}=\hat{\nabla}\phi({\bf r})(g_{11}-g_{22})/8\Omega$ defines the coupling strength to the vector potential.
By minimizing the energy functional $\mathcal{E}=\langle\Psi|(i\hbar\partial_{t}-\hat{H}_{\pm})|\Psi\rangle$ and dropping $\pm$ subscripts on $\rho_{\pm}$, $\Psi_{\pm}$ and ${\bf A}_{\pm}$, the mean-field Gross-Pitaevskii equation is $i\hbar\partial_t\Psi({\bf r},t)=\hat{H}_{\text{GP}}\Psi({\bf r},t)$ where 
\begin{equation}\label{dnlse}
\hat{H}_{\text{GP}}=\frac{1}{2m}(\hat{\bf p}-{\bf A})^2+\frac{1}{2}m\omega_{t}^{2}x^2+\frac{1}{2}m\omega_{\bot}^{2}{\bf r}_{\bot}^{2}+{\bf a}_{1}\cdot{\bf j}+g\rho
\end{equation}
with
\begin{equation}\label{jop}
{\bf j}=\frac{\hbar}{2mi}\left[\Psi\left(\hat{\nabla}+\frac{i}{\hbar}{\bf A}\right)\Psi^{*}-\Psi^{*}\left(\hat{\nabla}-\frac{i}{\hbar}{\bf A}\right)\Psi\right].
\end{equation}
Equation (\ref{dnlse}) along with the current operator (\ref{jop}) present a new opportunity with which to explore nonlinear phenomena with ultracold gases. It has already emerged how the transport properties of such a system exhibit chiral \cite{edmonds_2013a} effects in the continuum, as well as non-standard tunneling dynamics when studied in a discrete setting \cite{edmonds_2013b}.


{\it Elementary excitations.} We will consider a tightly confined anisotropic atomic cloud such that the dynamics is described by the one-dimensional form of equation (\ref{dnlse}) and (\ref{jop}). This approximation is valid when the axial trap frequency is much less than the radial frequency, $\omega_{t}\ll\omega_{\bot}$. Defining the phase of the laser as $\phi=kx$, and using the nonlinear transformation
\begin{equation}
\Psi(x,t)=\exp{\left(-i\frac{k}{2}x+\frac{ia_{1}}{\hbar}\int^{x}_{-\infty}dx'\rho(x',t)\right)}\Phi(x,t)
\end{equation}
the one-dimensional equation of motion becomes
\begin{equation}\label{eom_1d2}
i\hbar\frac{\partial\Phi}{\partial t}=\left[-\frac{\hbar^2}{2m}\partial_{x}^{2}+\frac{m\omega_{t}^{2}}{2}x^2-2a_{1}j(x)+g|\Phi|^2\right]\Phi,
\end{equation}
where $a_1=k(g_{11}-g_{22})/8\Omega S_t$ defines the current strength and $S_t$ is the transverse area of the cloud. Meanwhile, the gauge transformed current is given by $j(x)=(\hbar/m)\text{Im}(\Phi^{*}(x)\partial_{x}\Phi(x))$. The state of the superfluid component in this system is interesting, as the interacting gauge theory captured by Eq. \ref{eom_1d2} manifestly violates Galilean relativity. The existence of the superfluid state relies on Landau's criterion, which assumes that the physics is invariant in the co-moving frame. For our system, this is no longer the case. A direct consequence of this is that a Bose condensed system lacking Galilean relativity possesses two critical velocities \cite{zhu_2012,zheng_2013}. For the Galilean invariant system, these two velocities are equal. Here, we are interested in the collective excitations of the trapped chiral quantum gas, and expand the function $\Phi(x,t)$ in the form $\Phi(x,t)=\Phi_{0}(x,t)+\delta\Phi(x,t)$, where $\Phi_{0}(x,t)$ is the ground state and $\delta\Phi(x,t)$ is the excitations in question. The excitations take the form \cite{pethicksmith} $\delta\Phi(x,t)=\exp{(-i\mu t/\hbar)}\left(u(x)\exp{(-i\omega t)}-v^{*}(x)\exp{(i\omega^* t)}\right)$, where $u(x)$ and $v(x)$ are the mode functions, and $\omega$ defines the mode frequency. These definitions lead to the Bogoliubov-de Gennes equations
\begin{eqnarray}\label{bdgm1}
(\mathcal{L}_0+\mathcal{J}_0+2gn)u+(\mathcal{J}_0-gn)v&=&\hbar\omega u\label{bdgm1}\\
(\mathcal{J}_0+gn)u-(\mathcal{L}_0-\mathcal{J}_0+2gn)v&=&\hbar\omega v,\label{bdgm2}
\end{eqnarray}
where the ground state density is defined by $n=|\Phi_{0}|^2$, the term containing the kinetic energy operator and trapping potential is given as $\mathcal{L}_{0}=-\frac{\hbar^2}{2m}\partial_{x}^{2}+\frac{1}{2}m\omega_{t}^{2}x^2-\mu$, and $\mathcal{J}_0=i\frac{a_1\hbar}{m}[n\partial_x-\frac{1}{2}\partial_x n]$. Equation (\ref{eom_1d2}) can then be written in a more compact form as
\begin{equation}
\left\{\mathcal{L}_{0}-2a_{1}j(x)+gn\right\}\Phi=0.
\label{gpej}
\end{equation}
The Bogoliubov-de Gennes equations (\ref{bdgm1}) and (\ref{bdgm2}) along with the Gross-Pitaevskii equation, (\ref{gpej}) will provide the key to the stability properties of the chiral quantum gas. The stability properties of the excitations have been studied in \cite{leonhardt_2003} for instance, and often in the context of artificial event horizons using a hydrodynamic description of a system. In our case the situation is different, where the flow of the superfluid gives rise to an effective scattering length. For instance, we expect that the ground state does not couple to the current, in which case the physics is simply that of the standard one-dimensional Gross-Pitaevskii equation. The current operator will however couple to the collective excitations.

\begin{figure}
\includegraphics[width=8.5cm]{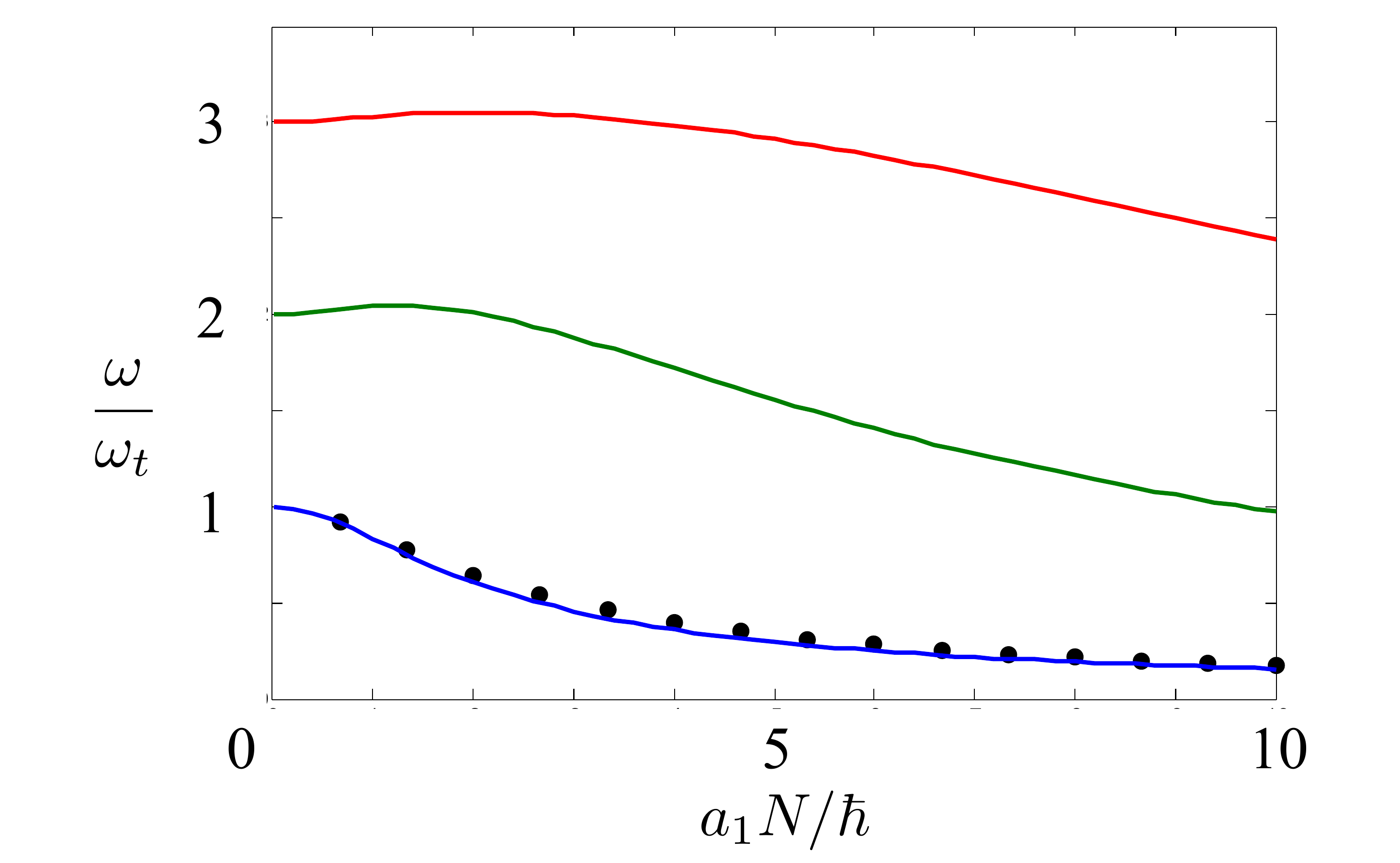}
\caption{The three lowest excitation frequencies from equation (\ref{bdgm1}) and (\ref{bdgm2}) plotted as a function of the current strength, $a_{1}N/\hbar$ in the limit $g=0$. The mode frequencies are all decreasing for strong $a_{1}N/\hbar$ values. The lowest excitation corresponds to the dipole mode which in the presence of a current nonlinearity depends on the $a_{1}N/\hbar$  parameter, in contrast to standard Bose-Einstein condensates. The dotted line indicates the results from the variational calculation.}
\label{fig1}
\end{figure}

There are two limits that are of interest. The first is  when $g=0$, which we refer to as the gaussian limit. The second is the Thomas-Fermi limit which is associated with the condition $gN\gg\hbar\omega_{t}a_{\text{ho}}$, where the harmonic length scale is defined as $a_{\text{ho}}=\sqrt{\hbar/m\omega_{t}}$. In the Thomas-Fermi limit the standard density nonlinearity is dominating over the current nonlinearity, and as such will not normally change the dynamics much compared to the standard spectrum of a trapped BEC. Figure \ref{fig1} shows the excitation frequencies in the gaussian limit, with $g=0$, which can be achieved if the sum of the scattering lengths are properly chosen. The ground state is then given by a gaussian profile, as the current operator in equation (\ref{eom_1d2}) will only affect the excitations. The frequencies of the lowest three excitations  are  plotted as a function of the dimensionless current strength $a_{1}N/\hbar$. 

One can see that there is a monotonic decrease in the frequency of the lowest mode, whereas the higher modes show for small values of the coupling strength $a_{1}N/\hbar$ an increase in frequency, but decrease in frequency for larger values of $a_{1}N/\hbar$.
The lowest excitation would in the normal case correspond to the dipole mode, which is an oscillating displacement of the centre of mass of the gas. This frequency does not depend on the interactions between the atoms, which is also referred to as the Kohn theorem. In the case of a current nonlinearity this is no longer the case. The dipole mode is strongly influenced by the current term. The underlying dynamics is governed by a coupling between the centre of mass and the width of the cloud, which is also not present in standard condensates. It should be noted that, in spin-orbit coupled condensates the Kohn theorem also does not hold, due to the coupling between the centre of mass and the internal spin degree of freedom \cite{ozawa2013,price_2013}.  

In the nonlinear regime (see Fig. 2 and 3), which means a strong coupling strength $a_{1}N/\hbar$ or sufficiently high momentum, we see a clear deviation from the sinusoidal dynamics given by the linearised spectrum above. In order to capture this dynamics we need to resort to a variational description. 

{\it Variational analysis.} 
In order to gain a better understanding of the dynamics, especially in the strongly nonlinear regime, a variational approach can be adopted which captures the main features of the dynamics. The motivation is two-fold. First we want to obtain an analytical expression for the frequency of the lowest mode by assuming small amplitude oscillations, hence small current nonlinearity. Secondly, the variational ansatz should provide an  accurate solution to the dynamics also in the intermediate nonlinear regime, meaning we have a well defined wave packet oscillating in the harmonic trap, but not captured by the Bogoliubov de Gennes treatment. For very strong nonlinearities we expect also the variational ansatz to break down, as shown below. 

We assume that the full time-dependent state of the system is known but different parameters such as the centre of mass coordinate and width of the condensate are allowed to vary. Hence, one is able to obtain equations of motion for these variational quantities which give insight into the dynamics of the system. This approach has been applied to the single component condensate \cite{garcia_1996}, and recently to spin-orbit coupled quantum gases \cite{zhang_2012,chen_2012} to explain the collective properties of the gas. Here, we adopt a similar approach to explore the dynamical properties of the many-body system. The normalized variational wave function is assumed to be a gaussian of the form
\begin{equation}
\Phi(x,t)=\left(\frac{N^2}{\pi\sigma_{x}(t)^2}\right)^{1/4}\exp{\left(-\frac{(x-x_{0}(t))^2}{2\sigma_{x}(t)^2}\right)}e^{i\mathcal{S}},
\label{ansatz}
\end{equation}   
where the variational parameters are the width $\sigma_{x}(t)$ and the centre of mass $x_{0}(t)$. The phase is given by 
\begin{equation}
\mathcal{S}=\frac{m}{\hbar}(\dot{x}_{0}x+\frac{1}{2}(x-x_0)^2\frac{\dot{\sigma}_{x}}{\sigma_{x}}).
\end{equation}
The resulting equations of motion for $x_{0}(t)$ and $\sigma_{x}(t)$ then become
\begin{eqnarray}
\ddot{x}_{0}+\omega^{2}_{t}x_{0}&=&\sqrt{\frac{2}{\pi}}\frac{a_{1}N}{m}\frac{\dot{\sigma}_{x}}{\sigma^{2}_{x}}\label{com}\\ 
\ddot{\sigma}_{x}+\omega^{2}_{t}\sigma_{x}&=&\frac{\hbar^2}{m^2\sigma^{3}_{x}}+\frac{N}{\sqrt{2\pi}}\frac{(g-4a_{1}\dot{x}_{0})}{m\sigma^{2}_{x}}.\label{width}
\end{eqnarray}
It can immediately be seen from these two equations that the dynamics  differ radically from the known behaviour of the single component gas. Equation (\ref{com}) shows how the centre of mass coordinate of the condensate no longer oscillates at the frequency of the trap, but instead has a nonlinear forcing term that depends on the width of the one dimensional cloud. Also, the equation of motion for the width (\ref{width}) has an extra term on the right hand side proportional to the velocity of the centre of mass. The small amplitude oscillations around the equilibrium points of equations (\ref{com}) and (\ref{width}) allow us to understand the low lying excitation frequencies of the condensate.


We choose $g=0$ in order to probe the effects of the current nonlinearity more clearly. By linearising equations (\ref{com}) and (\ref{width}) around $x_0=0$ and $\sigma=\sigma_0$ where $\sigma_0$ is the width of the Gaussian ground state, we expect to at least capture the main features of the lowest excitation provided that the fluctuations of the width and the amplitude of the centre of mass is much smaller than the width of the Gaussian ground state. The lowest excitation frequency after linearisation is readily given by
\begin{equation}
\omega=\frac{\omega_t}{\sqrt{2}}\sqrt{5+2\tilde a^2-\sqrt{4\tilde a^4+20\tilde a^2+9}}\label{freq}
\end{equation}
where $\tilde a=a_1N \sqrt{2}/(\sqrt{\pi}\hbar)$.  We see that the lowest excitation frequency decreases with increasing $\tilde a$. The frequency $\omega$ in Eq. (\ref{freq}) is found to be very close to the frequency calculated by the Bogoliubov-de Gennes equations shown in Fig. \ref{fig1}. For $\tilde a\ll 1$ we get  $\omega=\omega_t(1-a^2/3)$. For $\tilde a\gg 1$ the excitation frequency decreases as $\omega\sim 2\omega_t/|\tilde a|$. The equation of motions for the width and the centre of mass in eqs. (\ref{com}) and (\ref{width}) based on the variational calculation do not capture the full dynamics at strong current nonlinearity because we are restricted to only two modes of excitations, namely the dipole type mode and the breathing mode, and any linear combination of them, and do not take into account any asymmetry in the cloud which is to be expected from the current nonlinearity \cite{edmonds_2013a}. It does still describe the main features of the dynamics relatively well and can also capture the large amplitude limit where the centre of mass of the BEC is displaced much more than the width of the cloud.  

\begin{figure}
\includegraphics[width=8.5cm]{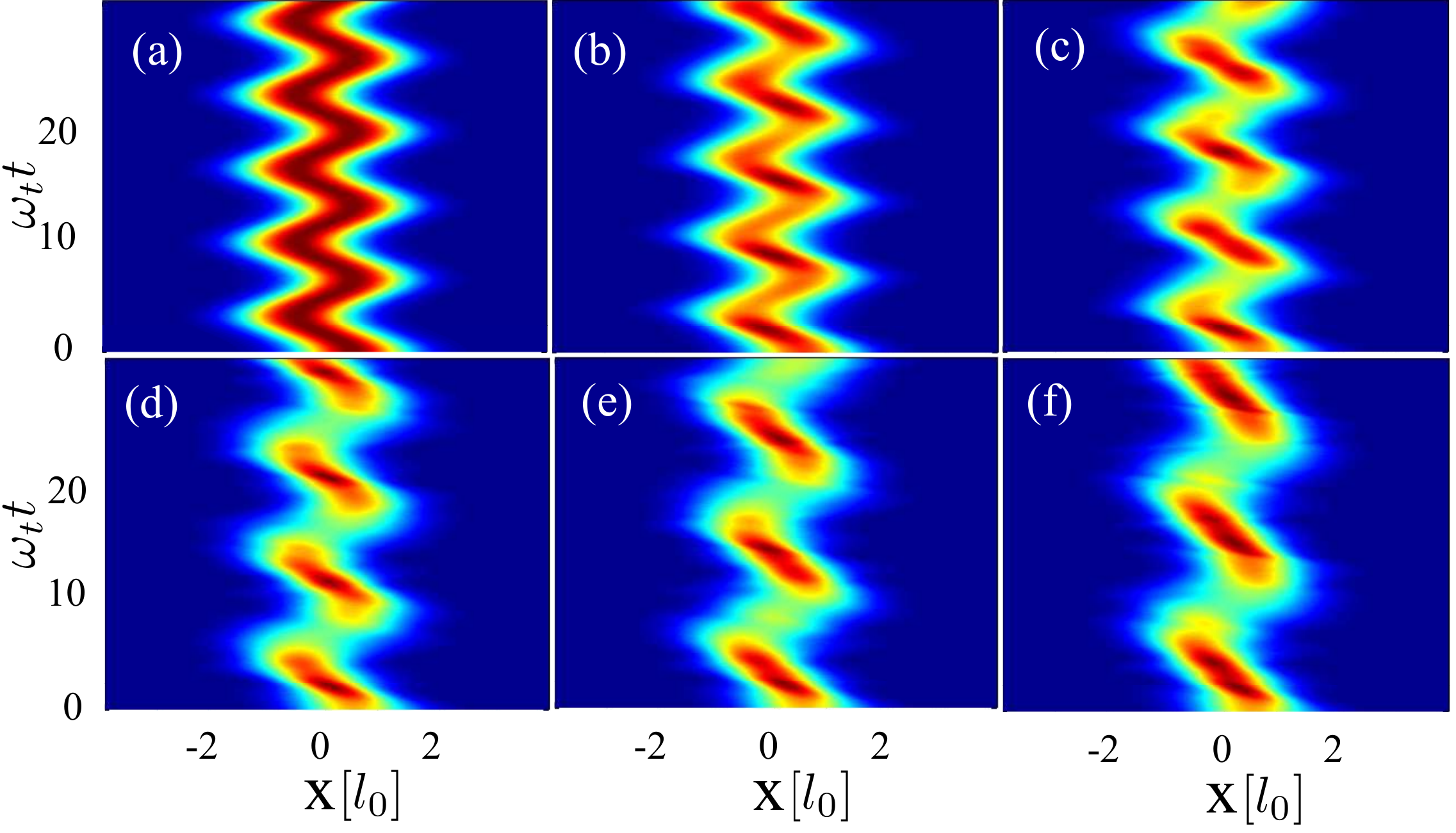}
\caption{(Color online) An initially displaced BEC centred at $x=1$ does not oscillate with the trap frequency once $a_1$ is increased. (a) $a_1$=0 corresponds to the standard harmonic oscillator dynamics. (b) $a_1N/\hbar=1$, (c)  $a_1N/\hbar=2$, (d) $a_1N/\hbar=3$, (e) $a_1N/\hbar=4$, (e) $a_1N/\hbar=5$.}\label{density1}
\end{figure}

\begin{figure}
\includegraphics[width=8.5cm]{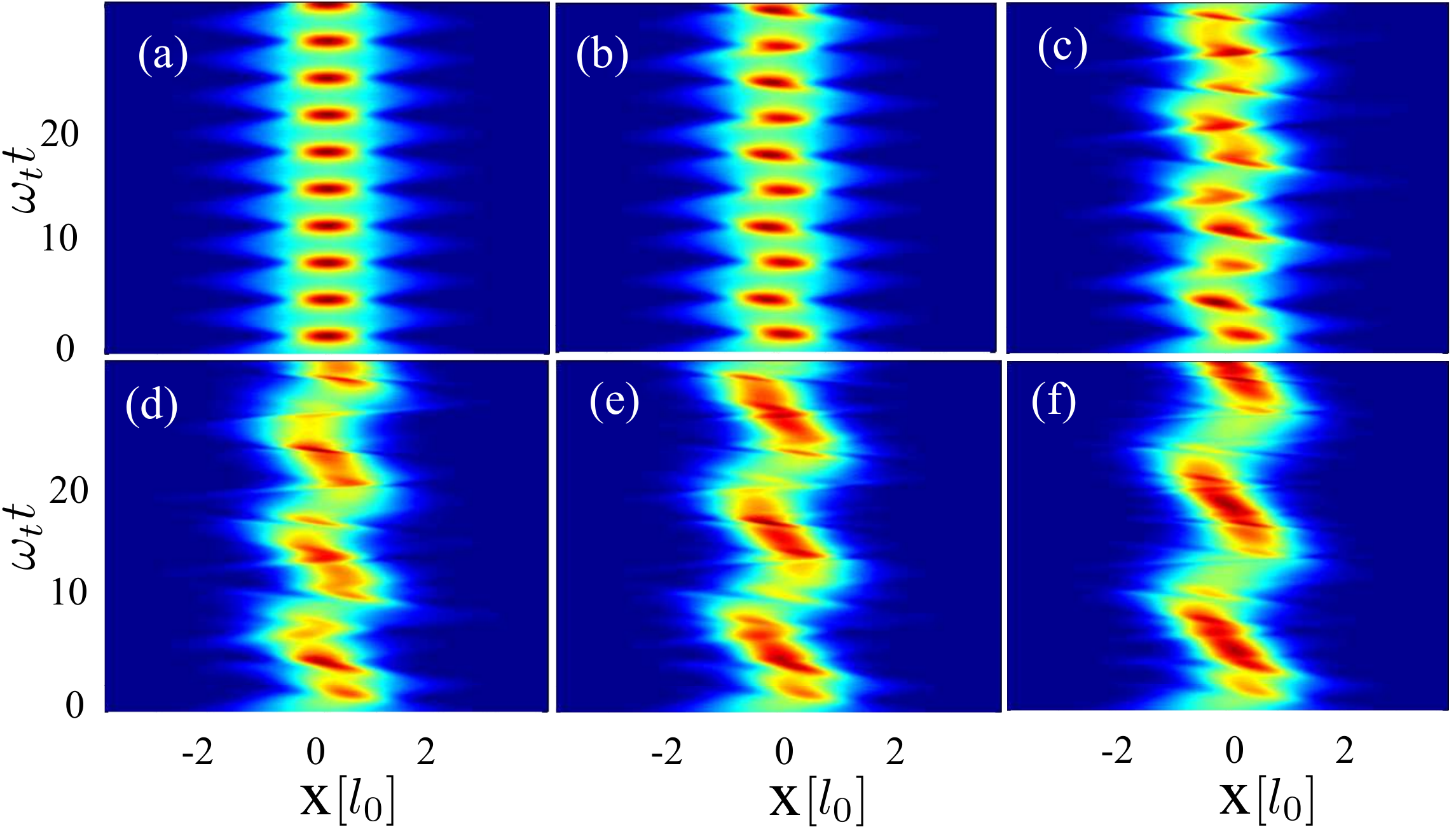}
\caption{(Color online) A BEC whose width at $t=0$ is larger than the Gaussian ground state width, centred at $x=0$, will start to couple to the dipole mode once $a_1$ is increased. (a) $a_1$=0 corresponds to the standard harmonic oscillator dynamics. (b) $a_1N/\hbar=1$, (c)  $a_1N/\hbar=2$, (d) $a_1N/\hbar=3$, (e) $a_1N/\hbar=4$, (e) $a_1N/\hbar=5$. For strong $a_1N/\hbar$ the dynamics relaxes to the same state found in Fig. \ref{density1} with a dipole frequency decreasing as $(a_1N/\hbar)^{-1}$ }\label{density2}
\end{figure}

A decreasing frequency with increasing current strength is also seen from the Bogoliubov-de Gennes spectrum, as shown in Fig. \ref{fig1}. In fact the linearised spectrum from the variational approach for the lowest mode, indicated by the dotted line in Fig. 1, agree remarkably well with the Bogoliubov-de Gennes spectrum. The Bogoliubov-de Gennes spectrum is based on a linearisation of the generalised Gross-Pitaevskii equation, and as such can describe the full spectrum, provided that the excitations do not deviate much from the ground state.  However, for strong current terms the chiral nature of the dynamics should also start to play a dominant role. A full numerical solution of the generalised Gross-Pitaevskii equation which contains a current nonlinearity confirms this, see Fig. \ref{density1} and  \ref{density2}. The decrease in excitation frequency is clearly seen in the dipole like excitation in Fig. \ref{density1}, but importantly,  both for an initial state corresponding to a translated wave packet or an initial width not equal to the ground state width, will result in the same dipole mode frequency. A type of non-equilibrium steady state situation is quickly reached corresponding to a dipole like mode which does not have the same frequency as the trap frequency. This would indicate that there is strong mode mixing taking place. With a current nonlinearity present the centre of mass motion is no longer decoupled from the other modes, which is the case for standard nonlinearities proportional to the density of the Bose-Einstein condensate. In the chiral system considered here, the breathing mode, which is normally a symmetric oscillation in the width of the cloud, is no longer present. For an initial gaussian state whose width is larger than the ground state width, we expect a breathing mode to be excited. But, the initial contraction of the cloud will be asymmetric due to the effective scattering length having a different sign for positive and negative $x$-values, at least initially. This will result in an asymmetry in the density of the cloud which will consequently couple to the dipole like mode. Therefore, any initial small symmetric perturbation of the gaussian state will result in an oscillation of the centre of mass, as seen in Fig. \ref{density2}. This dynamics is well captured by the variational ansatz described above.  

\begin{figure}
\includegraphics[width=8.70cm]{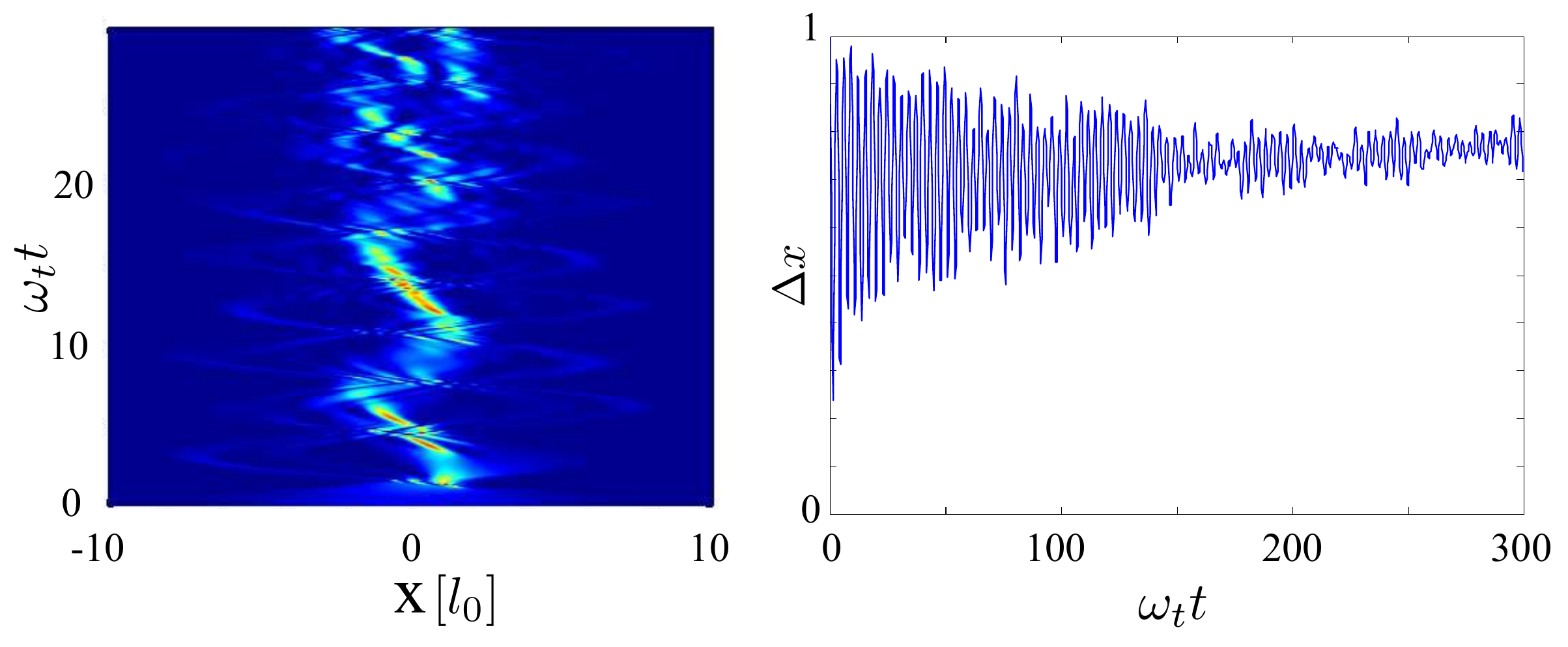}
\caption{(Color online) (a) For a broad initial state with $\sigma/\sigma_0=5$ the dynamics becomes irregular due to the strong influence of the current nonlinearity. (b) The normalised width $\Delta x(t)=\sqrt{\langle x^2\rangle-\langle x\rangle^2}/\Delta x(0)$ of the cloud as a function of time. No revivals can be seen. The $\Delta x$ relaxes towards a value smaller than 1 for long timescales.}\label{irr}
\end{figure}

{\it The strongly nonlinear regime.} For strong nonlinearities, $a_1\rho v \gtrsim \hbar\omega_0$, the dynamics becomes irregular, and cannot be described by a localised smooth wave packet any more. In Fig. \ref{irr} we illustrate this effect by considering a gaussian initial wave packet with a width $\sigma/\sigma_0=5$ and $a_1N/\hbar=5.0$. The wave packet quickly breaks up into irregular oscillations, with only a faint memory of the dipole mode in the dynamics. This leads us naturally to the question how irregular the dynamics is? 
In order to quantify the dynamics and to see if the system shows any kind of relaxation or revivals, we define the normalised width of the
cloud as $\Delta x(t)=\sqrt{\langle x^2\rangle-\langle x\rangle^2}/\Delta x(0)$. For a standard BEC with a nonlinearity proportional to the density of the cloud, we expect to see revivals, where the cloud oscillates symmetrically around the centre of the trap and with its characteristic frequency comes back to its initial state periodically. This scenario no longer holds with the current nonlinearity. Here we see no revivals. In addition there is also a degree of relaxation where the width $\Delta x$ approaches a value less than $1$. 

{\it Conclusions.} In this paper we have showed how a current nonlinearity stemming from an interacting gauge theory gives rise to unconventional superfluid dynamics, which manifests itself as a coupling between the centre of mass and the widths of the cloud. In the strongly coupled limit with a dominating current nonlinearity a dipole like mode is present where the oscillation frequency depends strongly on the strength of the current term. The chiral nature of the dynamics causes the effective scattering length of the gas to oscillate in sign and magnitude when the centre of mass of the BEC oscillates in the harmonic trap. Generally speaking, this effect might also allow for some intriguing applications related to the transport properties of the gas. The directional dependence of the nonlinearity could give rise to diode like behaviour of the quantum gas (see also \cite{ramezani_2010,larson_2011,haag_2014} for alternative schemes how to obtain diode like dynamics), which may provide a new mechanism for robust coherent control of superfluids.   

\begin{acknowledgements}
The authors would like to thank G. Juzeli\=unas, H. M. Price and L. Santos for useful discussions. M.J.E acknowledges support from an EPSRC Doctoral Prize Fellowship, M.V and P.\"O acknowledge support from EPSRC EP/J001392/1.
\end{acknowledgements} 


\end{document}